\documentclass[twocolumn,superscriptaddress,preprintnumbers,nofootinbib]{revtex4}
\usepackage{amsmath,amssymb}
\usepackage{graphicx}
\usepackage{epsfig,color}
\usepackage{bm}
\usepackage{verbatim}
\usepackage{amssymb}
\usepackage{hyperref}
\usepackage{ulem}
\usepackage{color}
\usepackage{bm}
\usepackage{soul}

\usepackage{soul,xcolor}

\begin{document}
\setstcolor{red}

\title{The $\Omega_{ c}$-puzzle solved by means of spectrum and  decay width predictions}


\author{E. Santopinto}
\affiliation{INFN, Sezione di Genova, Via Dodecaneso 33, 16146 Genova, Italy}
\author{ A. Giachino} 
\affiliation{INFN, Sezione di Genova, Via Dodecaneso 33, 16146 Genova, Italy}
\author{J. Ferretti}
\affiliation{Center for Theoretical Physics, Sloane Physics Laboratory, Yale University, New Haven, Connecticut 06520-8120, USA}
\author{H. Garc{\'i}a-Tecocoatzi}
\affiliation{Instituto de Ciencias Nucleares, Universidad Nacional Aut\'onoma de M\'exico, 04510 Ciudad de M\'exico, M\'exico} 
\author{M. A. Bedolla} 
\affiliation{INFN, Sezione di Genova, Via Dodecaneso 33, 16146 Genova, Italy}
\affiliation{Instituto de F\'isica y Matem\'aticas, Universidad Michoacana de San Nicol\'as de Hidalgo, Edificio C-3, Ciudad Universitaria, Morelia, Michoac\'an 58040, M\'exico}
\author{ R. Bijker} 
\affiliation{Instituto de Ciencias Nucleares, Universidad Nacional Aut\'onoma de M\'exico, 04510 Ciudad de M\'exico, M\'exico} 
\author{E. Ortiz-Pacheco}
\affiliation{Instituto de Ciencias Nucleares, Universidad Nacional Aut\'onoma de M\'exico, 04510 Ciudad de M\'exico, M\'exico} 
\begin{abstract}
The observation of five $\Omega_{ c}=ssc$ states by LHCb \cite{Aaij:2017nav} and the confirmation of four of them by  Belle \cite{Yelton:2017qxg}, may represent an important milestone in our understanding of the quark organization inside hadrons. By providing results for the spectrum of $\Omega_{ c}$ baryons and predictions for their $\Xi _{ c}^{+}K^{-}$ decay channels
within an harmonic oscillator based model, we suggest a possible solution to the $\Omega_{c}$ quantum number puzzle and  we extend our mass and decay width predictions to the $\Omega_b$ states.
Finally, we discuss why the set of $\Omega_{ c(b)}$ baryons is the most suitable environment to test the validity of three-quark and  quark-diquark  effective degrees of freedom.
\end{abstract}

\maketitle


\section{Introduction}
The discovery of new resonances always enriches the present experimental knowledge of the hadron zoo, and also provides essential information to explain the fundamental forces that govern nature. 
As the hadron mass patterns carry information on the way the quarks interact one another, they provide a means of gaining insight into the fundamental binding mechanism of matter at an elementary level.

In 2017, the LHCb Collaboration announced the observation of five narrow $\Omega_{ c}$ states in the $\Xi _{c}^{+}K^{-}$  decay channel \cite{Aaij:2017nav}: $\Omega_{ c}(3000)$, $\Omega_{ c}(3050)$, $\Omega_{ c}(3066)$, $\Omega_{ c}(3090)$ and $\Omega_{ c}(3119)$.
They also reported the observation of another structure around 3188 MeV,  the so-called $\Omega_{ c}(3188)$, though they did not have enough statistical significance to interpret it as a genuine resonance \cite{Aaij:2017nav}.
Later,  Belle  observed five resonant states in the $\Xi_c^{+} K^{-}$ invariant mass distribution and unambiguously confirmed four of the states announced by LHCb, $\Omega_{ c}(3000)$, $\Omega_{ c}(3050)$, $\Omega_{ c}(3066)$, and $\Omega_{ c}(3090)$, but no signal was found for the $\Omega_{ c}(3119)$ \cite{Yelton:2017qxg}. 
Belle also measured a signal excess at 3188 MeV,  corresponding to the $\Omega_{ c}(3188)$ state reported by LHCb 
\cite{Yelton:2017qxg}.
A comparison between the results reported by the two collaborations is displayed in Table \ref{tab:Table1}.
Here, it is shown that the $\Omega_{ c}(3188)$, even if not yet confirmed, was seen both by LHCb and   Belle, while, on the contrary, the $\Omega_{ c}(3119)$  was not observed by Belle.
\begin{table*}[htb]
\caption{Measured masses (in MeV) of the six resonances observed in $\Xi_c^{+}K^{-}$ decay channel (see text) according to the LHCb \cite{Aaij:2017nav} and  the Belle \cite{Yelton:2017qxg} collaborations in $ pp $ and $e^{+}e^{-}$ collisions, respectively.
}
\begin{tabular}
 {
  @{\hspace{0.1cm}}c@{\hspace{0.1cm}} | @{\hspace{0.1cm}}c@{\hspace{0.1cm}} |
  @{\hspace{0.1cm}}c@{\hspace{0.1cm}} | @{\hspace{0.1cm}}c@{\hspace{0.1cm}} |
  @{\hspace{0.1cm}}c@{\hspace{0.1cm}} | @{\hspace{0.1cm}}c@{\hspace{0.1cm}} | 
  @{\hspace{0.1cm}}c@{\hspace{0.1cm}}
}
\hline \hline
$\Omega_c$ excited state & 3000  & 3050 & 3066 & 3090 & 3119 & 3188  \\
Mass  (LHCb \cite{Aaij:2017nav})  & $3000.4\pm 0.2\pm 0.1$  & $3050.2\pm0.1\pm0.1$ & $3065.6\pm 0.1\pm0.3$ & $3090.2\pm 0.3\pm0.5$ & $3119\pm0.3\pm0.9$ &$3188\pm5\pm13$ \\
Mass (Belle \cite{Yelton:2017qxg})  & $3000.7\pm 1.0\pm0.2$  & $3050.2\pm0.4\pm0.2$ & $3064.9\pm0.6\pm0.2$ & $3089.3\pm1.2\pm0.2$ &-&$3199\pm9\pm4$ \\                 
\hline
\hline
\end{tabular}
\label{tab:Table1}
\end{table*}
It is also worth to mention that the LHCb collaboration has just announced the observation of a new bottom baryon, $\Xi_b(6227)^-$, in both $\Lambda_b^0 K^-$ and $\Xi^0_b \pi$ decay modes \cite{Aaij:2018yqz}, and of two bottom resonances, $\Sigma_b(6097)^\pm$, in the $\Lambda_b^0 \pi^\pm$ channels \cite{Aaij:2018tnn}. 

However, neither LHCb nor Belle were able to measure the $\Omega_c$ angular momenta and parities.
For this reason, several authors tried to provide different quantum number assignments for these states.
The current $\Omega_{c}$ puzzle consists in the discrepancy between the experimental results, reported by LHCb \cite{Aaij:2017nav} and  Belle~\cite{Yelton:2017qxg}, and the existing theoretical predictions~\cite{Karliner:2017kfm,Zhao:2017fov,Wang:2017hej,Padmanath:2017lng,Agaev:2017lip}. 
Indeed, for a given $\Omega_{ c}$ experimental state, more than one quantum number assignment was suggested~\cite{Karliner:2017kfm}. In particular, the $\Omega_{ c}(3119)$ was allocated to possibly be a $J^{P}=\frac{1}{2}^{+}$ or a $J^{P}=\frac{3}{2}^{+}$ state \cite{Wang:2017hej}, while the authors in Ref. \cite{Padmanath:2017lng} proposed a $J^{P}=\frac{5}{2}^{-}$ assignment.

From the previous discussion it comes out that, in the case of the $\Omega_{ c}(3119)$, not only the quantum number assignments are not univocal, but also the quark structure of the baryon is still unclear.
The issues we have to deal with are not restricted to the contrasts between the different interpretations provided in the previous studies, but are also related to the discrepancies 
on the quantum number assignments within a given study.
For example, in Ref. ~\cite{Agaev:2017lip} the authors provided different $J^P$ assignments for the $\Omega_c(3066)$ and  $\Omega_c(3090)$ based on mass and decay width estimates. 
Moreover, the nature of the $\Omega_c(3188)$ state is not addressed in these studies \cite{Karliner:2017kfm,Zhao:2017fov,Wang:2017hej,Padmanath:2017lng,Agaev:2017lip}.
These divergences between the theoretical interpretations created a puzzle which needs to be addressed urgently.

By estimating the contributions due to spin-orbit interactions, spin-, isospin- and flavour-dependent
interaction from the well-established charmed baryon mass spectrum, we reproduce quantitatively the spectrum of  the $\Omega_{c}$ states 
 within a harmonic oscillator hamiltonian plus a perturbation  term given by spin-orbit, isospin and flavour dependent  contributions (Secs. \ref{secIIA} and  \ref{secIIB}).
Based on our results, we  describe these five states as $P$-wave $\lambda$-excitations of the $ssc$ system; we also calculate their $\Xi _{ c}^{+}K^{-}$ decay widths (Sec. \ref{secIIC}).
Similarly to Refs.~\cite{Huang:2018wgr,Nieves:2017jjx,Debastiani:2017ewu}, we suggest  a molecular interpretation of the $\Omega_{ c}(3119)$ state, which was not observed by Belle.
Additionally, we extend our mass and decay width predictions to the $\Omega_b$ sector, which will be useful for future experimental searches.
\begin{figure}[htbp]
\begin{center}
\includegraphics[width=0.6\linewidth]{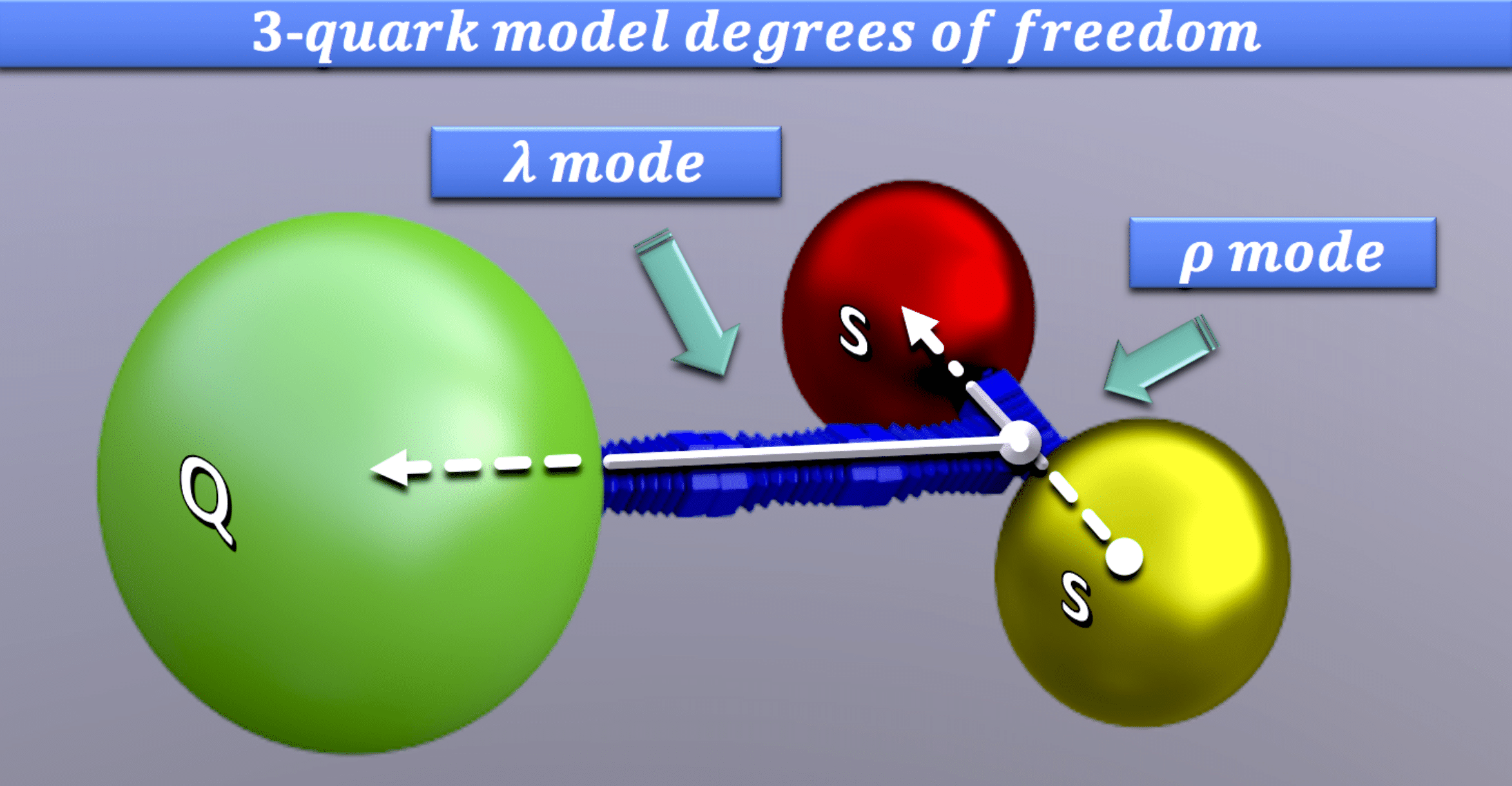}
\includegraphics[width=0.6\linewidth]{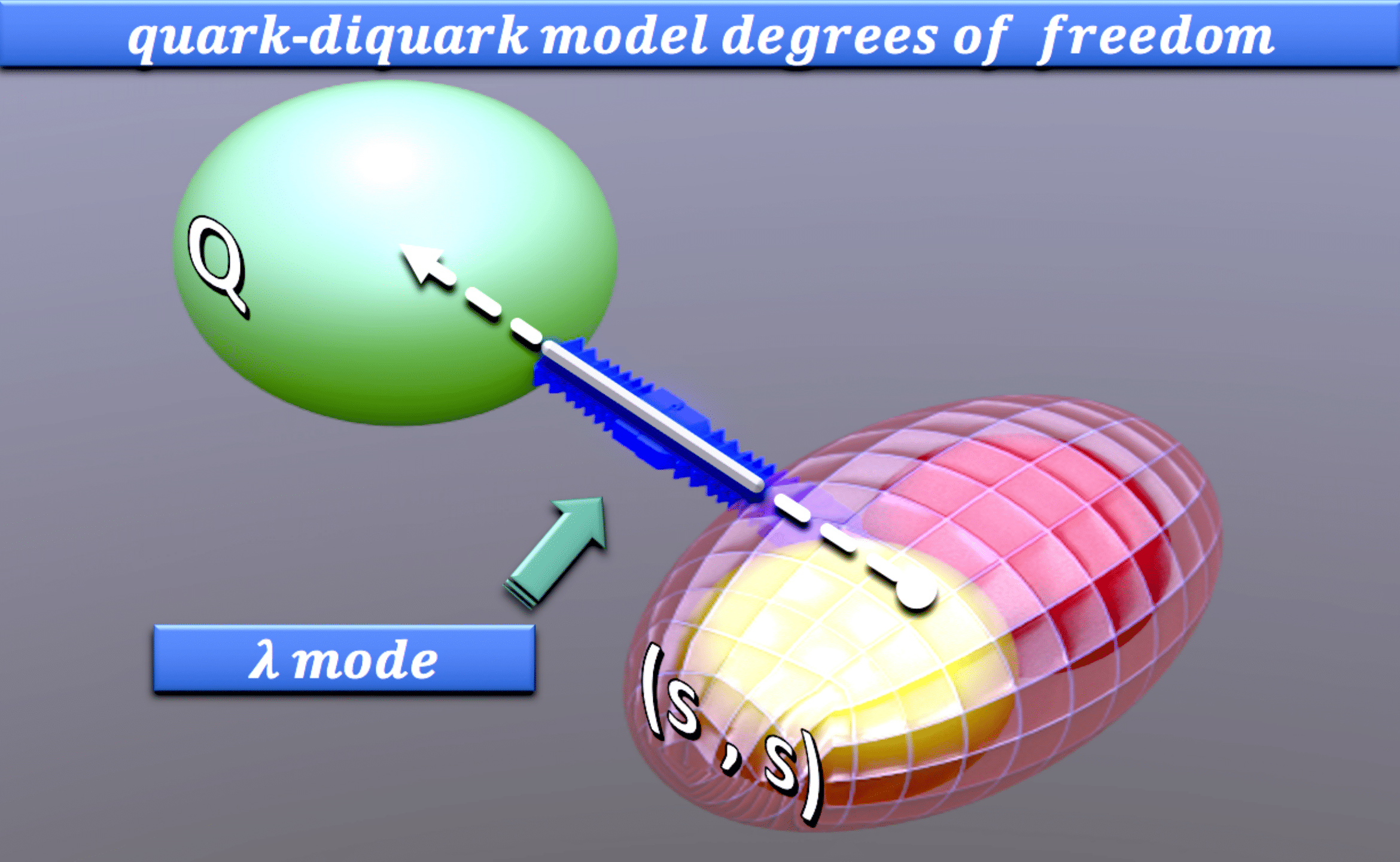}
\caption{Comparison between three-quark and quark-diquark baryon effective degrees of freedom. Upper  panel: three-quark picture with two excitation modes. Lower panel:  quark-diquark picture with one excitation mode.}
\label{comparison}
\end{center}
\end{figure}
Finally, we calculate  the mass splitting between the $\rho$- and $\lambda$-mode excitations of $\Omega_{c(b)}$ resonances (see Fig.~\ref{comparison} upper-pannel).
This calculation is  fundamental to access to inner heavy-light baryon structure,
as the presence or absence of $\rho$-mode excitations in the experimental spectrum
will be the key to discriminate between the three-quark (see Fig. \ref{comparison}  upper-pannel)  and the quark-diquark structures (see Fig. \ref{comparison}  lower-pannel),  as it will be discussed in Sec. \ref{qD}.

\section{Results}
\label{results}
\subsection{$S$- and $P$-wave $ssQ$ states.}
\label{secIIA}
The three-quark system ($ssQ$) Hamiltonian can be written in terms of two coordinates~\cite{Isgur:1978xj}, $\rho$ and $\lambda$, which encode the system spatial degrees of freedom (see Fig.~\ref{comparison}).
Let $m_{\rho}=m_s$ and $m_{\lambda}= \frac{3 m_s m_{Q}}{2m_s+m_{Q}}$ be the  $ssQ$ system reduced masses; then, the $\rho$- and $\lambda$-mode frequencies are $\omega_{\rho,\lambda}=\sqrt{\frac{3 K_Q}{m_{\rho,\lambda}}}$, which implies that in three equal-mass-quark baryons, in which $m_{\rho} = m_{\lambda}$, the $\lambda$- and $\rho$- orbital excitation modes are completely mixed together. 
By contrast, in heavy-light baryons, in which $m_{\rho} \ll m_{\lambda}$, the two excitation modes can be decoupled from each other as long as the light-heavy quark mass difference increases.

First of all, we construct the $ssc$ and $ssb$ ground and excited states to establish the quantum numbers of the five confirmed $\Omega_c$ states.
For simplicity, we use the compact notation $ssQ$ to denote them ($Q = c$ or $b$).
A single quark is described by its spin, flavor and color.
As a fermion, its spin is $S=\frac{1}{2}$, its flavor, spin-flavor and color representations are ${\bf {3}}_{\rm f}$, ${\bf 6}_{\rm sf}$, and ${\bf {3}}_{\rm c}$, respectively. 
An $ssQ$ state, $\left| ssQ, S_{\rho},S_{\rm tot}, l_{\rho}, l_{\lambda}, J \right\rangle$, is characterized by total angular momentum ${\bf J} = {\bf l}_{\rho}+{\bf l}_{\lambda}+  {\bf S}_{\rm tot}$, where ${\bf S}_{\rm tot}={\bf S}_{\rho}+\frac{\bf 1}{\bf 2}$.
In order to construct an $ssQ$ color singlet state, the light quarks must transform under SU$_{\rm c}$(3) as the anti-symmetric ${\bar {\bf 3}}_{\rm c}$ representation. 
The Pauli principle postulates that the wave function of identical fermions must be anti-symmetric for particle exchange. 
Thus, the $ss$  spin-flavor and orbital wave functions have the same permutation symmetry: symmetric spin-flavor in  S-wave, or antisymmetric spin-flavor  in antisymmetric $P$-wave.
Two equal flavour quarks are necessarily in the ${\bf {6}}_{\rm f}$ flavor-symmetric state.
 Thus, they are in an $S$-wave symmetric spin-triplet state, $S_{\rho}=1$, or in a $P$-wave antisymmetric spin-singlet state, $S_{\rho}=0$.

If $l_{\rho}=l_{\lambda}=0$, then $S_{\rho}=1$, and we find the two
ground states, $\Omega_Q$ and $\Omega_{Q}^{*}$:
$\left| ssQ,1,S_{\rm tot},0_{\rho},0_{\lambda},J \right\rangle$ with
$J=S_{\rm tot}=\frac{1}{2}$ and $\frac{3}{2}$, respectively.
If $l_{\rho}=0$ and $l_{\lambda}=1$, then $S_{\rho}=1$ and, by coupling
the spin and orbital angular momentum, we find five excited states:
$\left| ssQ,1,S_{\rm tot},0_{\rho},1_{\lambda},J \right\rangle$ with
$J=\frac{1}{2}$, $\frac{3}{2}$ for $S_{\rm tot}=\frac{1}{2}$,
and $J=\frac{1}{2}$, $\frac{3}{2}$, $\frac{5}{2}$ for $S_{\rm
tot}=\frac{3}{2}$, which we interpret as $\lambda$-mode excitations of the
$ssQ$ system.
On the other hand, if $l_{\rho}=1$ and $l_{\lambda}=0$, then $S_{\rho}=0$,
and we find two excited states
$\left| ssQ,0, \frac{1}{2},1_{\rho},0_{\lambda},J \right\rangle$ with
$J=\frac{1}{2}$, $\frac{3}{2}$ which we interpret as $\rho$- mode
excitations of the $ssQ$ system.

\subsection{Mass spectra of $\Omega_{Q}$ states}
\label{secIIB}
We employ a three-dimensional harmonic oscillator hamiltonian (h.o.)
plus  a perturbation  term given by spin-orbit, isospin and flavour dependent  contributions:
\begin{eqnarray}
	H = H_{\rm h.o.}+A\; {\bf S }^2 + B \; {\bf S} \cdot {\bf L} +E\;  \bm{I}^2+G \; {\bf C_2}(\mbox{SU(3)}_{\rm f}) ;
	\label{MassFormula}
\end{eqnarray}
here ${\bf S}, {\bm I}$ and ${\bf C_2}(\mbox{SU(3)}_{\rm f})$ are the spin, isospin and $\mbox{SU(3)}_{\rm f}$ Casimir operators, respectively, and  
\begin{eqnarray}
 H_{\rm h.o.} =\sum_{i=1}^3m_i + \frac{\mathbf{p}_{\rho}^2}{2 m_{\rho}} 
+ \frac{\mathbf{p}_{\lambda}^2}{2 m_{\lambda}} 
+\frac{1}{2} m_{\rho} \omega^2_{\rho} \boldsymbol{\rho}^2   
+\frac{1}{2}  m_{\lambda} \omega^2_{\lambda} \boldsymbol{\lambda}^2
	\nonumber \\
\label{Hho}
\end{eqnarray}
is the three-dimensional harmonic oscillator Hamiltonian, written in terms of the Jacobi coordinates, $ \boldsymbol{\rho}$ and $\boldsymbol{\lambda}$, and conjugated momenta, $\mathbf{p}_{\rho}$ and $ \mathbf{p}_{\lambda}$; its eigenvalues are $\displaystyle \sum_{i=1}^3m_i + \omega_{\rho}  n_{\rho} + \omega_{\lambda} n_{\lambda}$, where  $\omega_{\rho(\lambda)}=\sqrt{\frac{3K_Q}{m_{\rho(\lambda)}}}$, $ n_{\rho(\lambda)}= 2 k_{\rho(\lambda)}+l_{\rho(\lambda)}$, $k_{\rho(\lambda)}=0,1,...$, and $l_{\rho(\lambda)}=0,1,$ and so on.
We set the quark masses to reproduce the $\Omega_c(2695)$, $\Omega_c^{*}(2765)$, $\Xi_{cc}(3621)$ and $\Sigma_b(5814)$  ground state masses  \cite{Tanabashi:2018oca}: $m_q=295$ MeV, 
$m_{s}=450$ MeV, $m_c=1605$ MeV and $m_b=4920$ MeV;
the  spring constant  $K_c$ is set to reproduce  the mass difference between
$\Xi_c(2790)$, with $J^P=\frac{1}{2}^{-}$, and the $\Xi_c(2469)$ ground state:  $K_c=0.0328 $ 
GeV$^{3}$,
while  $K_b$ is set to reproduce  the mass difference between 
$\Lambda_b(5919)$, with $J^P=\frac{1}{2}^{-}$,  and the $\Lambda_b(5619)$ ground state:  $K_b=0.0235 $ GeV$^{3}$.
In order to calculate the mass difference between the $\rho$ and $\lambda$ orbital excitations of $ssQ$ states, we scale the h.o. frequency by the $\rho$ and $\lambda$  oscillator masses.
From the definition of  $m_{\rho}$  and $m_{\lambda}$, one finds $m_{\rho}=m_s=450$ MeV and $m_{\lambda}= \frac{3 m_s m_{c}}{2m_s+m_{c}}\simeq 865$ MeV for $\Omega_c$ states, and $m_{\lambda}= \frac{3 m_s m_{b}}{2m_s+m_{b}}\simeq 1141$ MeV for $\Omega_b$ states; the $\rho$- and $\lambda$-mode frequencies are $\omega_{\rho,\lambda}=\sqrt{\frac{3K_Q}{m_{\rho,\lambda}}}$.
Finally, the mass splitting parameters,  $A,B,E$ and $G$, calculated in the following, are reported in Table \ref{parameters}.


We estimate the mass splittings due to the spin-orbit, spin-, isospin- and flavor-dependent interactions from the well established charmed (bottom) baryon mass spectrum.
The spin-orbit interaction, which is mysteriously small in light baryons \cite{Capstick:1986bm,Ebert:2007nw}, turns out to be fundamental to describe the heavy-light baryon mass patterns, as it is clear from those of the recently observed  $\Omega_c$ states.
The spin-, isospin-, and flavour-dependent interactions are necessary to reproduce the masses of charmed baryon ground states, as observed in Ref. \cite{Santopinto:2016pkp}. 
By means of these estimates, we predict in a parameter-free procedure the spectrum of the $ssQ$ excited states constructed in the previous section.
The predicted masses of the $\lambda$- and $\rho$-orbital excitations of the $\Omega_c$ and $\Omega_b$ baryons are reported in Tables \ref{tab:widthsOmegac} and \ref{tab:mass prediction3}, respectively.
\begin{table*}[htbp]
\caption{Our $ssc$ state quantum number assignments (first column), predicted masses (second column) and strong decay widths (fourth column) are compared with the  experimental  masses (third column) and total decay widths (fifth column) \cite{Aaij:2017nav,Tanabashi:2018oca}. An $ssc$ state, $\left| ssc, S_{\rho}, S_{\rm tot}, l_{\rho}, l_{\lambda}, J \right\rangle$, is characterized by total angular momentum ${\bf J} = {\bf l}_{\rho}+{\bf l}_{\lambda} + {\bf S}_{\rm tot} $, where ${\bf S}_{\rm tot} = {\bf S}_{\rho}+\frac{1}{2}$. Our results are compatible with the experimental data, the predicted partial decay widths being lower than the total measured decay widths. The masses of states denoted by $\$$ are used as inputs, while all the other values are predictions; the partial decay widths denoted by $\dag$ and $\ddag$ are zero for phase space and for selection rules, respectively. }
\begin{tabular}{ccccc}
\hline
\hline
State & Predicted Mass & Experimental Mass & Predicted Width & Experimental Width\\
         & (MeV)                & (MeV)                      & $\Gamma(\Omega_c \rightarrow \Xi_c^{+}K^{-})$ (MeV) &  $\Gamma_{\rm tot}$  (MeV) \\
\hline
$\left| ssc,1, \frac{1}{2},0_{\rho},0_{\lambda}, \frac{1}{2} \right\rangle \equiv \Omega_{\rm c}(2695)^{\$}$   & 
$2702 \pm 12$ & $2695\pm2$ &  $\dag$ &  $<10^{-7}$ \\
$\left| ssc,1, \frac{3}{2},0_{\rho},0_{\lambda}, \frac{3}{2} \right\rangle \equiv \Omega_{\rm c}^{*}(2770)^{\$}$   & 
 $2767\pm13$ &   $2766\pm2$ & $\dag$  &  \\
$\left| ssc,1, \frac{1}{2},0_{\rho},1_{\lambda}, \frac{1}{2} \right\rangle \equiv \Omega_{\rm c}(3000)$   & 
$3016 \pm 9$ & $3000.4\pm0.2\pm0.1\pm0.3$ & 0.41  & $4.6\pm0.6\pm0.3$ \\
$\left| ssc,1, \frac{3}{2},0_{\rho},1_{\lambda},  \frac{1}{2} \right\rangle \equiv \Omega_{\rm c}(3050)$   &  $3045 \pm 13$   & $3050.2\pm0.1\pm0.1\pm0.3$ & 
0.42   & $0.8\pm0.2\pm0.1$\\
$\left| ssc,1,  \frac{1}{2},0_{\rho},1_{\lambda}, \frac{3}{2} \right\rangle  \equiv \Omega_{\rm c}(3066)$  &  $3052 \pm 15$ & $3065.6\pm0.1\pm0.3\pm0.3$ &
3.50     & $3.5\pm0.4\pm0.2$ \\
$\left| ssc,1, \frac{3}{2},0_{\rho},1_{\lambda}, \frac{3}{2} \right\rangle  \equiv \Omega_{\rm c}(3090)$  & 
 $3080\pm13$ & $3090.2\pm0.3\pm0.5\pm0.3$ &
0.75
  & $8.7\pm1.0\pm0.8$ \\
$\left| ssc,1, \frac{3}{2},0_{\rho},1_{\lambda}, \frac{5}{2} \right\rangle  \equiv \Omega_{\rm c}(3188)$  & 
$3140 \pm14$ &  $3188\pm5\pm13$ &
5.92
   & $60\pm26$   \\
 $\left| ssc,0, \frac{1}{2},1_{\rho},0_{\lambda}, \frac{1}{2} \right\rangle  $ &  $3146 \pm 12$ &  & $\ddag$ &  \\
  $\left| ssc,0, \frac{1}{2},1_{\rho},0_{\lambda}, \frac{3}{2} \right\rangle $ & $3182 \pm 12$ &  & $\ddag$ &  \\
\hline
\hline
\end{tabular}
\label{tab:widthsOmegac}
\end{table*}
In particular, Table \ref{tab:widthsOmegac} shows that we are able to reproduce quantitatively the mass spectra of the $\Omega_{c}$ states observed both by LHCb and  Belle; the latter are reported in Table \ref{tab:Table1}.
\begin{table*}[htbp]
\caption{Our $ssb$ state quantum number assignments (first column), predicted masses (second column) and strong decay widths (fourth column) are compared with the  experimental  masses (third column) and total decay widths (fifth column) \cite{Tanabashi:2018oca}. An $ssb$ state, $\left| ssb, S_{\rho}, S_{\rm tot}, l_{\rho}, l_{\lambda}, J \right\rangle$, is characterized by total angular momentum ${\bf J} = {\bf l}_{\rho}+{\bf l}_{\lambda} + {\bf S}_{\rm tot} $, where ${\bf S}_{\rm tot} = {\bf S}_{\rho}+\frac{1}{2}$. The partial decay widths denoted by $\dag$ and $\ddag$ are zero for phase space and for selection rules, respectively.
}
\centering
\begin{tabular}{ccccc}
\hline
\hline
State & Predicted Mass &Experimental Mass  &  Predicted Width & Experimental Widths   \\
&(MeV) & (MeV)& $\Gamma(\Omega_{\rm b} \rightarrow \Xi_b^{0}K^{-})$ (MeV) & $\Gamma_{\rm tot}$  (MeV) \\
\hline
$\left| ssb,1, \frac{1}{2},0_{\rho},0_{\lambda}, \frac{1}{2} \right\rangle \equiv \Omega_b $ & $6061\pm 15$&  $6046 \pm2 $&
 $\dag$ & $<10^{-9}$   \\
$\left| ssb,1, \frac{3}{2},0_{\rho},0_{\lambda}, \frac{3}{2} \right\rangle $  &  $6082\pm 20$   & &
$\dag$  &  \\
 $\left| ssb,1, \frac{1}{2},0_{\rho},1_{\lambda}, \frac{1}{2} \right\rangle $ &  $6305\pm 15$   &&
0.22  &  \\
$\left| ssb,1,  \frac{3}{2},0_{\rho},1_{\lambda},   \frac{1}{2} \right\rangle  $  &     $6317\pm 19$ &&
0.43  &      \\
  $\left| ssb,1, \frac{1}{2},0_{\rho},1_{\lambda},  \frac{3}{2} \right\rangle $ &      $6313\pm 15$
  && 1.49 &  \\
    $\left| ssb,1, \frac{3}{2},0_{\rho},1_{\lambda}, \frac{3}{2} \right\rangle $  &     $6325\pm 19$   &&
    0.32  &    \\
      $\left| ssb,1, \frac{3}{2},0_{\rho},1_{\lambda}, \frac{5}{2} \right\rangle $  & $6338\pm 20$   &&    
      2.11 &  \\
 $\left| ssb,0, \frac{1}{2},1_{\rho},0_{\lambda}, \frac{1}{2} \right\rangle  $ &   $6452\pm 15$    && $\ddag$ &    \\
  $\left| ssb,0, \frac{1}{2},1_{\rho},0_{\lambda}, \frac{3}{2} \right\rangle $ &  $6460\pm 15$   &&  $\ddag$ & \\
\hline
\hline
\end{tabular}
\label{tab:mass prediction3}
\end{table*}

\begin{figure}[htbp]
\caption{$\Omega_c$ mass spectra and tentative quantum number assignments. The theoretical predictions (red dots) are compared with the experimental results by LHCb ~\cite{Aaij:2017nav} (blue line), Belle \cite{Yelton:2017qxg} (violet line) and Particle Data Group (black lines) \cite{Tanabashi:2018oca}. Except the $\Omega_c(3188)$ case, the experimental error for the other states is too small to be appreciated in this energy scale. The spin-$\frac{1}{2}$ and -$\frac{3}{2}$ ground-state masses, $\Omega_c(2695)$
and $\Omega_c^{*}(2770)$ are indicated with $\dagger$  because are inputs while all the others are predictions.}
\begin{center}
\includegraphics[width=8.4cm]{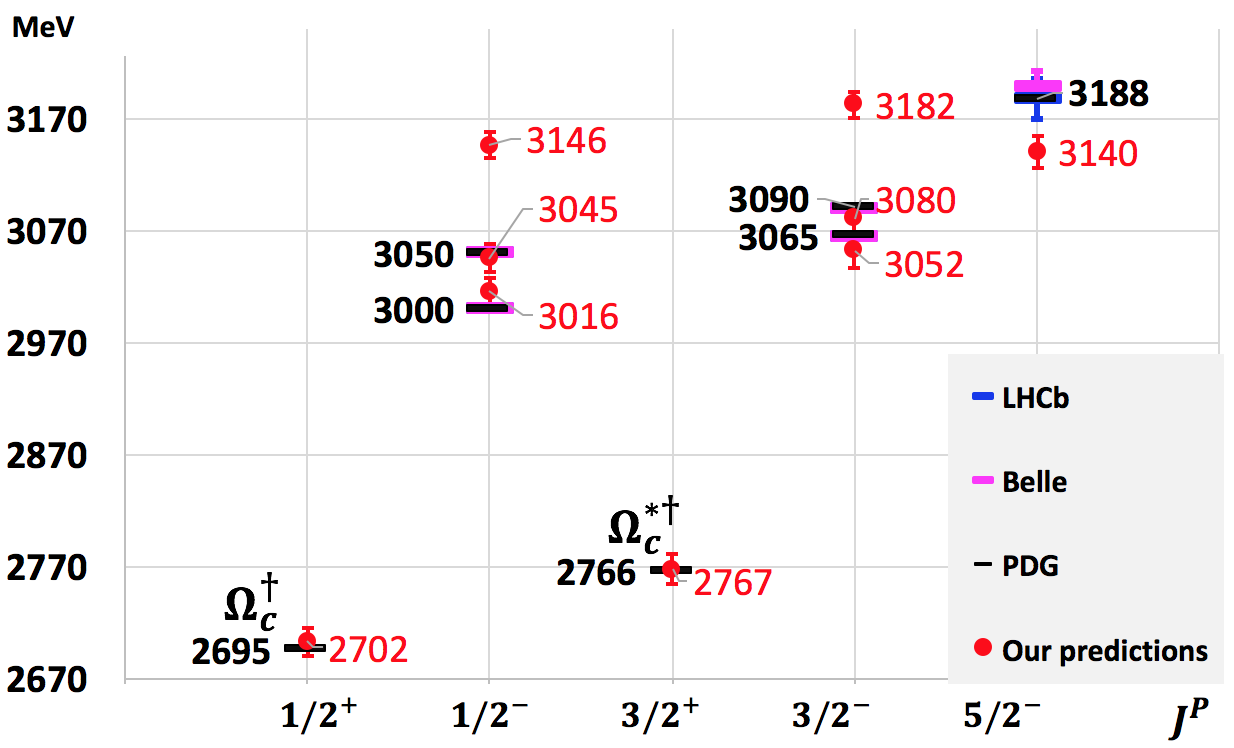}
\label{spectrum1}
\end{center}
\end{figure}  

We estimate the energy splitting due to the spin-spin interaction from the (isospin-averaged) mass difference between $\Sigma_{c}^*(2520)$  and $\Sigma_{c}(2453)$. This value ($65 \pm 8$ MeV) agrees with the mass difference between $\Omega_{c}$ (2695) and $\Omega_{c}^*$ (2770), a value close to $71$ MeV. 
As a consequence, the spin-spin mass splitting between two  orbitally excited states characterized by the same flavor configuration but different spins, specifically $S_{\rm tot}=\frac{1}{2}$ and $S_{\rm tot}=\frac{3}{2}$, is around $65$ MeV plus corrections due the spin-orbit contribution which can be calculated, for example, from the $\Lambda_{ c}(2595)$-$\Lambda_{ c}(2625)$ mass difference. 
According to the quark model, $\Lambda_{ c}(2595)$ and $\Lambda_{ c}(2625)$ are the charmed counterparts of $\Lambda_{ }(1405)$ and $\Lambda_{ }(1520)$, respectively; their spin-parities are $\frac{1}{2}^{-}$ and $\frac{3}{2}^{-}$, and their mass difference, about 36 MeV, is due to spin-orbit effects.  

\begin{figure}[htbp]
\caption{$\Omega_b$ mass spectrum predictions (red dots) and $\Omega_b$ ground-state experimental mass (black line)  \cite{Tanabashi:2018oca}.
The experimental error on the $\Omega_b(6046)$ state,  2 MeV,  is too small to be appreciated in this energy scale.} 
\begin{center}
\includegraphics[width=8.3 cm]{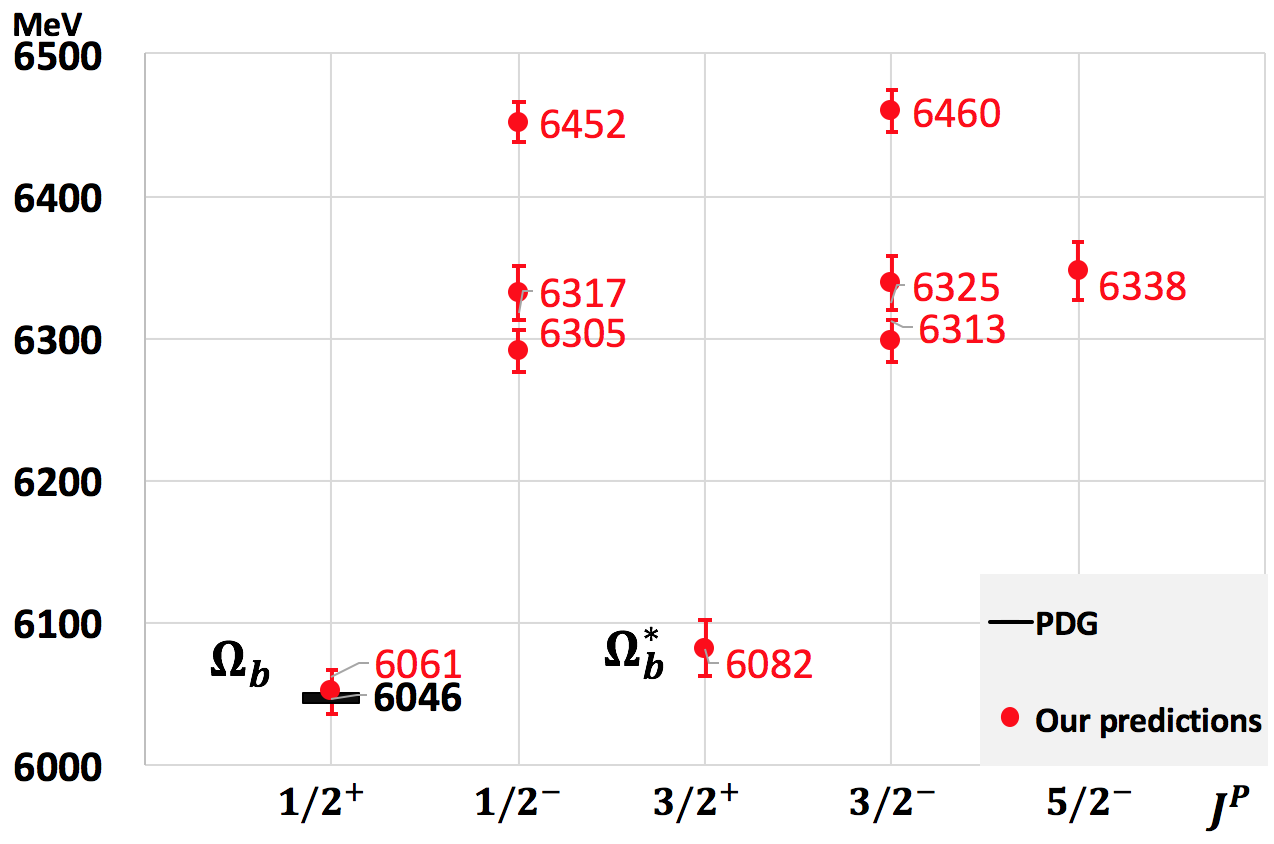}
\label{spectrum2}
\end{center}
\end{figure}

In conclusion, by taking into account the spin-spin and the spin-orbit contributions,  the mass difference between the lowest $\Omega_c$ excitation, $\left| ssc,1, \frac{1}{2},0_{\rho},1_{\lambda},\frac{1}{2} \right\rangle  \equiv \Omega_c(3000)$,  and $\left| ssc,1, \frac{3}{2},0_{\rho},1_{\lambda},\frac{1}{2} \right\rangle$
is about $65-36\simeq 30 $ MeV, and so we identify the  
$\left| ssc,1, \frac{3}{2},0_{\rho},1_{\lambda},\frac{1}{2} \right\rangle$ with the observed $\Omega_c(3050)$ (see Fig.  \ref{spectrum1}  and Table \ref{tab:widthsOmegac} ).
In the bottom sector, the energy splitting due to the spin-spin interaction through the (isospin-averaged) mass difference between $\Sigma_{b}^*$ and $\Sigma_{b}$ is $20 \pm 7$ MeV.
In such a way, we expect a mass difference between the two $S$-wave ground states, $\Omega_{b}^{*}$ and $\Omega_{b}^{}$, close to $20 \pm 7$ MeV. Hence, we suggest the experimentalists to look for a 
$\Omega_{b}^{*}$ resonance with a mass of about 6082 MeV, as we can see in Figure \ref{spectrum2} and Table 
\ref{tab:mass prediction3}.

We estimate that the mass of $\left| ssc,1, \frac{1}{2},0_{\rho},1_{\lambda}, \frac{3}{2} \right\rangle$ is related to the previous spin-orbit splitting. We obtain a value of $3052 \pm 15$ MeV,  which is compatible with the mass of the $\Omega _{ c}^{} (3066)$  within the  experimental error. Thus, we identify the $\left| ssc,1, \frac{1}{2},0_{\rho},1_{\lambda}, \frac{3}{2} \right\rangle$ state with the 
$\Omega_{ c}(3066)$ resonance.
Through the estimation of orbital, spin-spin and spin-orbit interactions, we estimate the $\left| ssc,1, \frac{3}{2},0_{\rho},1_{\lambda}, \frac{3}{2} \right\rangle$ and $\left| ssc,1, \frac{3}{2},0_{\rho},1_{\lambda}, \frac{5}{2} \right\rangle$ mass values as 3080 $\pm $13 MeV and 3140 $\pm$14, respectively. 
Thence, we propose the following assignments: $\left| ssc,1, \frac{3}{2},0_{\rho},1_{\lambda}, \frac{3}{2} \right\rangle$ $\to$ $\Omega_c$(3090) and $ \left| ssc,1, \frac{3}{2},0_{\rho},1_{\lambda},\right.$ $\left. \frac{5}{2} \right\rangle$ $\to$ $\Omega_c$(3188).

In the bottom sector, the mass splitting due to the spin-orbit interaction between $\Lambda_{b}(5912)$ and $\Lambda_{b}(5920)$ is 8 MeV  and we estimated previously that the 
spin-spin splitting is $20 \pm 7$ MeV.
Thus, we  interpret the predicted $ \Omega_b (6305)$, $\Omega_b (6313)$, $\Omega_b (6317)$, $\Omega_b (6325)$ and $ \Omega_b (6338)$ states, reported in Table \ref{tab:mass prediction3}, as the bottom counterparts of the $\Omega _{ c}(3000)$,  $\Omega_c(3066)$, $\Omega_c(3050)$, $\Omega_c(3090)$ and $\Omega_c(3188)$, respectively.
We observe that, unlike the charm sector,  in the bottom sector  the state $\left| ssb,1, \frac{3}{2},0_{\rho},1_{\lambda}, \frac{1}{2} \right\rangle$ is heavier than the state $\left| ssb,\right.$ $\left.1, \frac{1}{2},0_{\rho},1_{\lambda}, \frac{3}{2} \right\rangle$: this is due to the fact that in the charm sector the spin-orbit contribution is lesser than the spin-spin one, while in the bottom sector the situation is the opposite (see Table \ref{parameters}).



In the charm sector, the mass splitting  due to the flavor-dependent interaction can be estimated from the mass difference between $\Xi_c$ and  $\Xi_c^{'}$, whose isospin-averaged masses are 2469.37 MeV and 2578.1 MeV, respectively; this leads to a value of 109 MeV, approximately.
The bottom partner of $\Xi_c$ and  $\Xi_c^{'}$ are $\Xi_b$ and  $\Xi_b^{'}$, with masses 5793.2 MeV and 5935.02 MeV, respectively. 
Therefore, in the bottom sector the flavor-dependent interaction gives a contribution of about 142 MeV, 
 which is more than $30\%$ larger than in the charm sector. 
The mass difference between the lightest charmed ground states, $\Sigma_c$ and  $\Lambda_c$, is related to the different isospin and flavor structures of the light quark multiplets: $\Lambda_c$ is an isospin-singlet state belonging to an SU(3)$_{\rm f}$ flavor anti-triplet,  while $\Sigma_c$ is an isospin-triplet state belonging to an SU(3)$_{\rm f}$ flavor sextet. In the bottom sector, the isospin-flavor contribution to the baryon masses can be calculated from the mass difference between $\Sigma_b$ and $\Lambda_b$.

\begin{table}[htbp]
\caption{Values of the parameter reported in Eq. (\ref{MassFormula}) with the corresponding uncertainties
expressed in MeV.}
\centering
\begin{tabular}{ccc}
\hline
\hline
State & $A$ & $B$ \\
\hline
charm   &  $21.54 \pm0.37$   &      $23.91\pm0.31$   \\
bottom  & $6.73\pm1.63$   &   $5.15\pm0.33$    \\
\hline
State & $E$ & $G$ \\
\hline
charm   & $30.34 \pm0.23$   &$54.37 \pm0.58$  \\
bottom  & $26.00\pm1.80$ &  $70.91\pm0.49$ \\
\hline
\hline
\end{tabular}
\label{parameters}
\end{table}
We summarize all our proposed quantum number assignments for both $\Omega_c$ and $\Omega_b$ states in Figs.~\ref{spectrum1} and \ref{spectrum2}, respectively. In the charm sector, we find a good agreement between the mass pattern predicted  for the spectrum and the experimental data:
in particular, with the exception of the lightest and the heaviest resonant states, $\Omega_c(3000)$
and $\Omega_c(3188)$, respectively, also the absolute mass predictions are in agreement within the 
experimental error, which is very small (less than 1 MeV).

\subsection{Decay widths of $ssQ$ states}
\label{secIIC}
In the following, we  compute the strong decays of $ssQ$ baryons in $sqQ - K$ ($q = u,d$) final states by means of the $^3P_0$ model ~\cite{Micu:1968mk,LeYaouanc:1972vsx,LeYaouanc:1988fx,Roberts:1997kq} (see \ref{app2}).
The $^3P_0$ model parameters are  the harmonic oscillator frequency of $K$ meson wave function,
  $\omega_{c}=0.46 $ GeV \cite{Close:2005se}, the pair creation strength,  $\gamma_0 =17.25 $,
 set to reproduce the $\Omega_c(3066)$ experimental decay width, and
 the baryon $\rho$- and $\lambda$-mode frequencies, $\omega_{\rho,\lambda}=\sqrt{3K_Q/m_{\rho,\lambda}}$, which  are
the same as that calculated to predict the mass spectrum.
\begin{figure}[htbp]
\caption{\textit{Adapted from Fig. 2 of  Ref.~\cite{Aaij:2017nav}, APS copyright.}  \newline
Proposed spin- and parity-assignments for the $\Omega_{ c} = css$ excited states  reported by the LHCb Collaboration and later observed by Belle: $\Omega_{ c}^{}(3000)$, $\Omega_{ c}^{}(3050)$, $\Omega_{ c}^{}(3066)$,  $\Omega_{ c}^{}(3090)$, and $\Omega_{ c}^{}(3188)$. We interpret $\Omega_{ c}(3119)$ as a $\Xi_c^{*}K$ molecule.}
\begin{center}
\includegraphics[width=7cm]{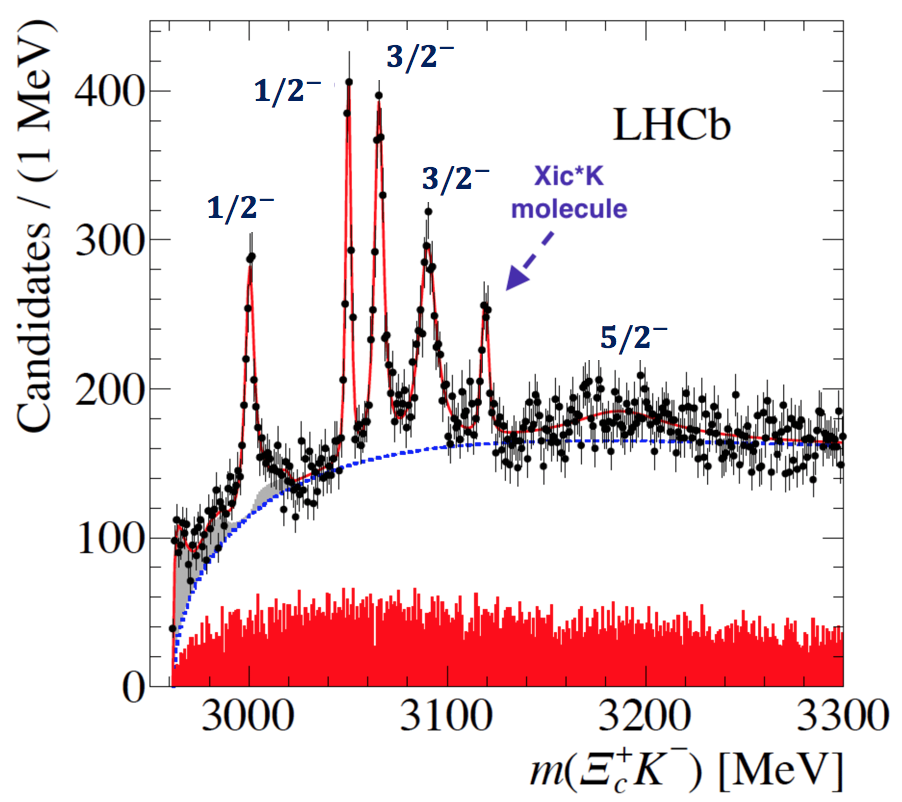}
\label{fig:proposedassignment}
\end{center}
\end{figure}

Tables \ref{tab:widthsOmegac} and \ref{tab:mass prediction3} report our $\Omega_c \to \Xi_c^{+}K^{-}$ and $\Omega_b\to \Xi_b^{0}K^{-}$ predicted decay widths.
The $ \Xi_c^{+}K^{-}$ decay channel is where the $\Omega_c$ states were observed by LHCb and Belle.  
Both the $\Xi_c^{+}K^{-}$ branching ratios and the quantum numbers of the $\Omega_c$'s are unknown; we only have experimental informations on their total widths, $\Gamma_{\rm tot}$. Thus, our predictions have to satisfy the constraint: $\Gamma(\Omega_c \to \Xi_c^{+}K^{-})\leq \Gamma_{\rm tot}$.
In light of this, we state that our strong decay width results, based both on our mass estimates and quantum number assignments, are compatible with the present experimental data.
In particular, the $\lambda$-mode decay widths of the $\Omega_c$ states are in the order 1 MeV,
while $\Xi_{\rm c}^{+}K^{-}$ decay of the two $\rho$-excitations, $\left| ssc,0, \frac{1}{2},1_{\rho},0_{\lambda}, \frac{1}{2} \right\rangle$ and $\left| ssc,0, \frac{1}{2},1_{\rho},0_{\lambda}, \frac{3}{2} \right\rangle$, is forbidden by spin conservation.
Similar considerations can be applied to the decay widths of $\rho$-mode $\Omega_b$ states.

In conclusion, in addition to our mass estimates, also the $^3P_0$ model results suggest that the five $\Omega_{ c}$ resonances, $\Omega_{ c}^{}(3000)$, $\Omega_{ c}^{}(3050)$, $\Omega_{ c}^{}(3066)$, $\Omega_{ c}^{}(3090)$, and $\Omega_{ c}^{}(3188)$, could be interpreted as $ssc$ ground-state $P$-wave $\lambda$-excitations. 
In principle, both the $\Omega_{ c}^{}(3090)$ and $\Omega_{ c}^{}(3119)$ resonances observed by LHCb are compatible with the properties (mass and decay width) of the $\left| ssc, \frac{3}{2},1_{\lambda},\frac{3}{2} \right\rangle$ theoretical state.
As Belle could neither confirm nor deny the existence of the $\Omega_{ c}^{}(3119)$, given the low significance of its results for the previous state ($0.4 \sigma$), we prefer to: 1) Assign $\left| ssc, \frac{3}{2},1_{\lambda},\frac{3}{2} \right\rangle$ to the $\Omega_{ c}^{}(3090)$; 2) Interpret the $\Omega_{ c}^{}(3119)$ as a $\Xi_c^{*}K$  bound state \cite{Huang:2018wgr,Debastiani:2017ewu,Nieves:2017jjx}, the $\Omega_{ c}^{}(3119)$ lying 22 MeV below the $\Xi_c^{*}K$ threshold. See Fig.~\ref{fig:proposedassignment}.

\section{Comparison between the three-quark and  quark-diquark structures}
\label{qD}
In the light baryon sector, in which all the constituent quarks have roughly the same mass, the two oscillators, $\rho$- and $\lambda$, have approximately the same frequency, $\omega_{\rho}\simeq \omega_{\lambda}$, 
which implies that the $\lambda$ and $\rho$ excitations are degenerate in the mass spectrum. 
By contrast, in the heavy-light baryons,  in which $m_{\rho} \ll m_{\lambda}$, the two excitation modes can be decoupled from each other as long as the light-heavy quark mass difference increases; 
specifically $\omega_{\rho}-\omega_{\lambda}\simeq 130$ MeV for the $\Omega_c$ states
and $\omega_{\rho}-\omega_{\lambda}\simeq 150$ MeV for the $\Omega_b$ states.
Thus,  the heavy-light baryon sector is the most suitable environment 
to test what are   the correct effective spacial degrees of freedom for reproducing 
the mass spectra, as the presence or absence of $\rho$-mode excitations in the  spectrum
will be the key to discriminate between the three-quark and the quark-diquark structures (see Fig. \ref{comparison}):  if  the predicted four $\rho$ excitations, $\Omega_c(3146)$,$\Omega_c(3182)$,$\Omega_b(6452)$, and $\Omega_b(6460)$ will not be observed 
then the other states possess a quark-diquark structure.
\indent Finally, we observe that in the case of a  quark-diquark-picture experimental confirmation
our model Hamiltonian,  Eq. \ref{MassFormula}, still holds, as the quark-diquark h.o. Hamiltonian, is the  limit of the three-quark h.o. Hamiltonian, Eq. \ref{Hho}, in which the $\rho$ coordinate is frozen\footnote{In this limit, we recover the expression of the well known two-body h.o. Hamiltonian in the quark-diquark centre of mass frame, $H_{\rm ho}=2m_D+m_Q+\frac{p_{r}^2}{2 \mu}+\frac{1}{2} \mu_{\rm r} \omega^2_{\rm r} {\bf r}^2$, where $\mu_{\rm r}=\frac{m_D m_Q}{m_D+m_Q}$ is the reduced mass of the quark-diquark system, ${\bf p}_{r}=\frac{m_Q {\bf p}_D-m_D {\bf p}_Q}{m_D+m_Q}$ the quark-diquark relative momentum, ${\bf r}={\bf r}_{D}-{\bf r}_Q$ the quark-diquark relative coordinate, and $\omega_{\rm r}$ the angular frequency of the quark-diquark oscillator. 
\\The two-body ($QD$) h.o. Hamiltonian is recovered from the three-body ($ssQ$) one when the $\rho$-oscillator collapses provided that one performs a rescaling of the spring constant of the oscillator, $K_Q$, due to different definitions of the the angular frequencies, $\omega_\lambda = \sqrt{\frac{3 K_Q}{m_\lambda}}$ and $\omega_{\rm r} = \sqrt{\frac{K_Q}{\mu_{\rm r}}}$, and the reduced masses, $m_\lambda$ and $\mu_{\rm r}$, of the $\lambda$ and QD oscillators.}:
\begin{eqnarray}
 H_{\rm h.o.} &=&m_{D}+m_Q +
\frac{\mathbf{p}_{\lambda}^2}{2 m_{\lambda}}  
+\frac{1}{2}  m_{\lambda} \omega^2_{\lambda} \boldsymbol{\lambda}^2.
\label{Hho2}
\end{eqnarray}
Here $m_D=2m_s$ is the diquark mass. Indeed, the  mass spectrum predicted with this definition of
$H_{\rm h.o.}$ is the same as that reported  in Figs.  ~\ref{spectrum1} and \ref{spectrum2}, but without the 
frozen $\rho$ excitations.
We also observe that: I) If the quark-diquark scenario turns out to be the correct one, the heavy quark symmetry, predicted by the heavy quark effective theory, HQET, in the heavy-light meson sector, can be extended to the heavy-quark-light-diquark baryon one; II) The suppression of spin-spin interactions, as we move from the charmed to the bottom sector, is another manifestation of the validity of the HQET in the heavy-light baryon sector.

\section{Discussion}
We calculate the $\Omega_{c(b)}$'s masses and $\Xi _{ c(b)}^{+}K^{-}$ strong decay amplitudes. 
By means of these mass and decay width predictions, we propose an univocal assignment to the five $\Omega_{c}$ states observed both by LHCb \cite{Aaij:2017nav} and Belle \cite{Yelton:2017qxg}: 
\begin{eqnarray}
&\hspace{1.5cm}\left| ssc, 1, {\scriptstyle \frac{1}{2}},0_{\rho},1_{\lambda},{\scriptstyle \frac{1}{2}} \right\rangle \to  \Omega_{ c}(3000)\,,&\\
&\hspace{1.5cm}\left| ssc,1, {\scriptstyle \frac{3}{2}},0_{\rho},1_{\lambda},{\scriptstyle \frac{1}{2}} \right\rangle \to  \Omega_{ c}(3050)\,,&\\
&\hspace{1.5cm}\left| ssc, 1,{\scriptstyle \frac{1}{2}},0_{\rho},1_{\lambda},{\scriptstyle \frac{3}{2}} \right\rangle  \to  \Omega _{ c}(3066)\,,&\\
&\hspace{1.5cm}\left| ssc, 1, {\scriptstyle \frac{3}{2}}, 0_{\rho},1_{\lambda},{\scriptstyle \frac{3}{2}} \right\rangle \to  \Omega _{ c}(3090)\,,&\\
&\hspace{1.5cm}\left| ssc, 1, {\scriptstyle \frac{3}{2}}, 0_{\rho},1_{\lambda},{\scriptstyle \frac{5}{2}} \right\rangle  \to \Omega_{ c}(3188)\,.&
\end{eqnarray}
The latter was completely ignored in other studies~\cite{Karliner:2017kfm,Zhao:2017fov,Wang:2017hej,Padmanath:2017lng,Agaev:2017lip}. 
In principle, both the $\Omega_{ c}(3119)$ and $\Omega_{ c}(3090)$ could be assigned to the
 $|ssc,1,\frac{3}{2},0_{\rho},1_{\lambda},\frac{3}{2}\rangle$ state. 
However, as Belle could neither confirm nor deny the existence of the $\Omega_{ c}(3119)$, we 
 prefer the $\Omega_{ c}(3119)$ interpretation as a $\Xi_c^{*} K$ meson-baryon molecule and assign the $\Omega_{ c}(3090)$ to the $\left| ssc, 1, \frac{3}{2},\right.$ $\left. 0_{\rho},1_{\lambda},\frac{3}{2} \right\rangle \to \Omega _{ c}(3090) $ state,  providing a consistent solution to the $\Omega_{ c}$ puzzle. 
\\
We calculate the mass splitting between the $\rho$- and $\lambda$-mode excitations of the $\Omega_{c(b)}$ resonances. 
 This large mass splitting, that  we predict  to be  greater than 150 MeV, is  fundamental to access to inner heavy-light baryon structure.
If the $\rho$-excitations  in the predicted mass region will not be observed 
in the future, then the three-quark model effective degrees of freedom for the heavy-light baryons 
will be  ruled out, supporting the Heavy Quark Effective Theory (HQET) picture of the heavy-light baryons
described as heavy quark-light diquark systems.
If the HQET is valid for the heavy-light baryons, the heavy quark symmetry, predicted by the HQET in the heavy-light meson sector, can be extended  to the heavy-quark-light-diquark baryon sector,
opening the way to new future theoretical applications.

\appendix

\section{$^3P_0$ Decay model}
\label{app2}
The $^3P_0$ is an effective model to compute the open-flavor strong decays of hadrons in the quark model formalism~\cite{Micu:1968mk,LeYaouanc:1972vsx,LeYaouanc:1988fx,Roberts:1997kq}. 
In this model, a hadron decay takes place in its rest frame and proceeds via the creation of an additional $q \bar q$ pair with vacuum quantum numbers, i.e. $J^{PC} = 0^{++}$.
We label the initial baryon- and final baryon- and meson-states as $A$, $B$ and $C$, respectively. The final baryon-meson state $BC$ is characterized by a relative orbital angular momentum $\ell$ between $B$ and $C$ and a total angular momentum ${\bf J} = {\bf J}_B + {\bf J}_C + {\bm \ell}$.
The decay widths can be calculated as \cite{Micu:1968mk,LeYaouanc:1972vsx,Ferretti:2015ada}
\begin{equation}
	\Gamma = \frac{2 \pi \gamma_0^2}{2J_{A}+1} \mbox{ } \Phi_{A \rightarrow BC}(q_0)\sum_{M_{J_A},M_{J_B}} \big|\mathcal{M}^{M_{J_A},M_{J_B}}\big|^2 \mbox{ }.  \nonumber
        \label{gamma}
\end{equation}
Here, $\mathcal{M}^{M_{J_A},M_{J_B}}$ is the $A \rightarrow BC$ amplitude which, for simplicity, is usually expressed in terms of hadron harmonic-oscillator wave functions, $\gamma_0$ is the dimensionless pair-creation strength. 
$q_0$ is the relative momentum between $B$ and $C$, and the coefficient $\Phi_{A \rightarrow BC}(q_0)$ is the relativistic phase space factor~\cite{Ferretti:2015ada},
\begin{eqnarray}
	\label{eqn:rel-PSF}
	\Phi_{A \rightarrow BC}(q_0) = 4 \pi q_0 \frac{E_B(q_0) E_C(q_0)}{M_A}  \mbox{ }, \nonumber\\
	\mbox{with}\;\;E_{B,C} = \sqrt{M_{B,C}^2 + q_0^2}.  \nonumber
\end{eqnarray}


\section*{Acknowledgments}
The authors acknowledge financial support from CONACyT, M\'exico (postdoctoral fellowship for M.A.~Bedolla), the INFN Sezione di Genova, the Instituto de Fisica y Matematicas of Universidad Michoacana de San Nicolas de Hidalgo,  the US Department of Energy Grant No. DE-FG-02-91ER-40608 (J. Ferretti),
and the Instituto de Ciencias Nucleares, Universidad Nacional Aut\'onoma de M\'exico.

\end{document}